\pdfoutput=1

\documentclass[conference]{IEEEtran}
\IEEEoverridecommandlockouts

\usepackage{cite}
\usepackage{amsmath,amssymb,amsfonts}
\usepackage{graphicx}
\usepackage{textcomp}
\usepackage{xcolor}
\usepackage{booktabs}
\usepackage{siunitx}
\usepackage{tikz}
\usetikzlibrary{positioning,arrows.meta,fit,backgrounds,calc}
\definecolor{urlcol}{HTML}{1A4E8A}
\usepackage[colorlinks=true, urlcolor=urlcol, linkcolor=black,
            citecolor=black, breaklinks=true]{hyperref}
\usepackage{url}
\DeclareSIUnit{\rearth}{\ensuremath{R_\oplus}}
\DeclareSIUnit{\rsun}{\ensuremath{R_\odot}}
\DeclareSIUnit{\ppm}{ppm}

\definecolor{teal}{HTML}{0C5E4E}
\definecolor{gold}{HTML}{9A8B45}
\definecolor{brick}{HTML}{C0392B}

\newcommand{\bt}{\mathrm{BTJD}}
\newcommand{\bj}{\mathrm{BJD}}

\tikzset{
  blk/.style={draw=teal, line width=0.7pt, rounded corners=2pt, fill=white,
              inner sep=3.2pt, align=center, font=\scriptsize, minimum height=8.5mm,
              text width=20mm},
  lbl/.style={draw=gold, line width=0.7pt, rounded corners=2pt, fill=gold!8,
              inner sep=3.2pt, align=center, font=\scriptsize, minimum height=8.5mm,
              text width=20mm},
  bad/.style={draw=brick, line width=0.7pt, dashed, rounded corners=2pt, fill=brick!5,
              inner sep=3.2pt, align=center, font=\scriptsize, minimum height=7mm,
              text width=20mm},
  fl/.style={-{Latex[length=4pt,width=3pt]}, line width=0.6pt, draw=teal},
  flb/.style={-{Latex[length=4pt,width=3pt]}, line width=0.6pt, draw=brick, dashed},
}

\begin{document}

\title{Astro-Hunters: Machine Learning for Exoplanet\\
Transit Detection in TESS Photometry}

\author{\IEEEauthorblockN{Fatimah Emad Eldin}
\IEEEauthorblockA{\textit{Cairo University}\\
12422024441586@pg.cu.edu.eg}
\thanks{\raggedright Trained weights, gold-standard labels and the labelled
corpus: \url{https://huggingface.co/FatimahEmadEldin/Astro-Hunters-Transit-Detector}.
Code reproducing every table and figure:
\url{https://github.com/astral-fate/Astro-Hunters-Model-traning}.
Hosted inference:
\url{https://huggingface.co/spaces/FatimahEmadEldin/Astro-Hunters}.
See Section~\ref{sec:availability}.\par}}

\maketitle

\begin{abstract}
Finding planets beyond the Solar System has become an exercise in automated data
analysis: space-based photometry surveys return far more stellar light curves
than can be inspected by eye, and machine learning is increasingly asked to
recognise the faint, periodic dimming of a transiting planet. Whether such a
detector works turns on a design choice that is seldom reported --- how its
training labels were made. No catalogue disposes the individual measurements a
per-cadence detector must classify, so the annotation must be derived; this
paper asks what that costs.
We present Astro-Hunters, an end-to-end pipeline over TESS two-minute
photometry: retrieval, detrending, seven sliding-window statistics per cadence,
and gradient-boosted classification. These components are deliberately
conventional. What is new is the treatment of label provenance as an
experimental variable; a corpus of \num{189279} cadences from twelve confirmed
hosts, annotated from published ephemerides rather than from the photometry
itself; and a physical bound on what the task permits.
Six classifier families are compared under a star-disjoint protocol.
Three results follow. Holding features, model and protocol fixed,
precision--recall performance spans a factor of \num{29} across label sources
against \num{1.8} across architectures: labels from an isolation forest fitted
to the classifier's own features give an apparent AUC of \num{0.9915} that
measures circularity, and an unconverted transit epoch drives performance to
chance, whereas correct annotation gives AUC \num{0.788} at \num{5.3} times the
prevalence baseline. Second, the ceiling is observational, not architectural:
a median single-cadence signal-to-noise ratio of \num{2.10} caps per-cadence
AUC at \num{0.932}. Third, a Box Least Squares baseline recovers eight of twelve
orbital periods from a single sector. Phase-folding, not classifier capacity, is
what makes the transit signal accessible.
\end{abstract}


\section{Introduction}
\label{sec:intro}

\subsection{Transits and the photometric signal}

A planet whose orbital plane is aligned closely enough with the line of sight
passes periodically in front of its host star. During such a transit the planet
occults a small fraction of the stellar disc, and the total flux received from
the system falls. For a planet of radius $R_p$ crossing a star of radius $R_*$,
the fractional depth of the resulting dip is, to first order,
\begin{equation}
\delta \;\simeq\; \left(\frac{R_p}{R_*}\right)^{\!2},
\label{eq:depth}
\end{equation}
so that a Jupiter-sized planet transiting a solar-type star produces a decrement
near one per cent, while an Earth-sized planet produces one near one part in
$10^4$. The transit repeats with the orbital period $P$ and lasts, for a central
crossing of a circular orbit, approximately
\begin{equation}
T_{14} \;\simeq\; \frac{P}{\pi}\,\frac{R_*}{a},
\label{eq:duration}
\end{equation}
with $a$ the orbital semi-major axis. The fraction of an orbit spent in transit,
the \emph{duty cycle} $T_{14}/P$, is therefore small: for the systems studied
here it ranges from \num{1.96}\,\% to \num{8.73}\,\%.

\begin{figure*}[t]
\centering
\includegraphics[width=\textwidth]{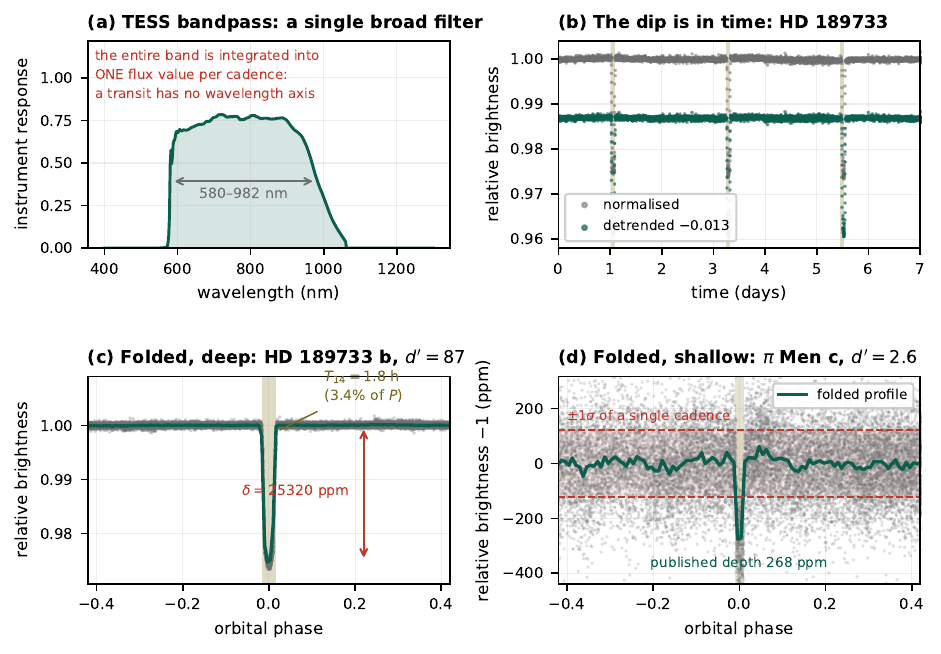}
\caption{The transit signal, and why it has no wavelength axis.
(a) The measured TESS instrument response~\cite{ricker2015}. The mission
integrates the whole \SIrange{580}{982}{\nano\metre} band into a single flux
value per cadence, so the observation carries no spectral dimension; depth
against wavelength is transmission spectroscopy, a different measurement.
(b) Seven days of HD\,189733, normalised and detrended, with the published
transit windows shaded: the dip is in \emph{time}, repeating at the orbital
period.
(c) The same star phase-folded, with transit depth $\delta$ and duration
$T_{14}$ marked. The transit occupies \num{3.4}\,\% of the orbit.
(d) $\pi$\,Men\,c folded, plotted in parts per million. The published
\SI{268}{\ppm} depth lies within the $\pm 1\sigma$ band of a single cadence
(shaded red), so no individual measurement can resolve it; folding recovers it.
Ephemerides throughout are from the NASA Exoplanet Archive, converted
$\bj \rightarrow \bt$; nothing is fitted to the photometry.}
\label{fig:transit}
\end{figure*}

Transit photometry is a broadband measurement, and this point is worth making
precisely because it determines the shape of the data. TESS observes through a
single wide filter whose measured response is shown in
Fig.~\ref{fig:transit}(a): it spans \SIrange{580}{982}{\nano\metre} at half
maximum, centred on the Cousins $I$ band and extended redward to favour cool,
small stars whose planets produce deeper signals for a given planetary
radius~\cite{ricker2015}. Every photon collected across that entire band is
summed into \emph{one} brightness value per cadence. The observation therefore
has no wavelength axis: a transit is not a dip in wavelength but a dip in flux
against \emph{time}, as Fig.~\ref{fig:transit}(b) shows. To first order it is
achromatic --- a geometric occultation, not a spectral feature.

Wavelength dependence does exist in transit measurements, but it is a different
observation. In transmission spectroscopy the depth $\delta(\lambda)$ varies by
tens to hundreds of parts per million across wavelength because the planetary
atmosphere is opaque at some wavelengths and transparent at others, and
resolving that requires a spectrograph. TESS cannot make that measurement; it
records a single integrated flux. This is why the transit-detection literature,
and this paper, operate on one-dimensional flux time series --- the
\emph{light curve} --- rather than on spectra.

Fig.~\ref{fig:transit}(c) and (d) show the signal this paper is trying to
detect, in the two regimes that matter. For HD\,189733\,b the dip is
\SI{25320}{\ppm} deep against a per-cadence scatter of \SI{292}{\ppm}, and is
plainly visible in a single event. For $\pi$\,Men\,c the published depth is
\SI{268}{\ppm} against a scatter of \SI{122}{\ppm}: no individual measurement
resolves it, and the transit becomes unambiguous only after folding on the
orbital period stacks the in-transit cadences. That contrast is the subject of
Section~\ref{sec:res-snr}.

\subsection{The scale of the archive}

TESS surveys the sky in sectors of approximately \num{27} days, observing a
preselected set of targets at a two-minute cadence and the full field at longer
cadence~\cite{ricker2015}. A three-hour transit sampled every two minutes is
described by roughly ninety measurements, enough to resolve the characteristic
flat-bottomed, sharply-ingressing profile. The resulting volume is far beyond
manual inspection: querying a single sector returns close to \num{13000}
distinct targets. Automated detection is consequently not a convenience but a
requirement, and it is the reason machine learning has been applied to this
problem for the better part of a decade.

\subsection{What was built}

Astro-Hunters is a complete detection pipeline together with the evaluation
apparatus needed to establish what it can and cannot do. Its stages are:
retrieval of Science Processing Operations Center light curves from the Mikulski
Archive for Space Telescopes via \texttt{lightkurve}; stitching of multiple
sectors, removal of invalid cadences, median normalisation and median-filter
detrending; extraction of seven local statistics over a sliding window of
$\pm 32$ cadences; and a gradient-boosted tree classifier assigning each cadence
a transit probability. Trained models are served through a public inference
endpoint, described in Section~\ref{sec:arch}.

The paper then subjects the pipeline's central modelling assumption --- that a
cadence is a sufficient unit of decision --- to a quantitative test, and
establishes the physical bound within which any classifier over these inputs
must operate.

\subsection{Contributions}

\begin{enumerate}
\item A controlled demonstration that, holding features, model and evaluation
      protocol fixed, the choice of label source moves the area under the
      precision--recall curve from \num{0.0298} to \num{0.8584}, a factor of
      \num{29}, while six model architectures span only a factor of \num{1.8}
      (Sections~\ref{sec:res-models} and~\ref{sec:res-ablation}). Neither
      extreme is a valid measurement of transit detection.
\item Identification and reproduction of two distinct label defects: circular
      labelling by an unsupervised detector fitted to the classifier's own
      feature matrix, and a time-system error in which archival transit
      midpoints in $\bj$ are applied to light curves expressed in $\bt$
      (Section~\ref{sec:res-ablation}).
\item A physical bound on the task itself. We measure the single-cadence
      signal-to-noise ratio for every host and show that the median value of
      \num{2.10} caps the per-cadence Bayes-optimal AUC at \num{0.932}, so that
      the corrected model, at \num{0.788}, and the best of six architectures, at
      \num{0.814}, both sit close to the information ceiling
      (Section~\ref{sec:res-snr}).
\item A classical Box Least Squares baseline on identical data recovering eight
      of twelve periods from a single sector, establishing that the signal is
      present and that the limitation is the decision unit, not the data
      (Section~\ref{sec:res-bls}).
\item Gold-standard per-cadence labels for twelve systems, constructed from
      published ephemerides over all transiting planets, released with the code.
\end{enumerate}

\section{Related Work}
\label{sec:related}

\subsection{Classical detection}

The standard detection primitive is a periodic box search. Box Least Squares
(BLS) folds the light curve at each of a grid of trial periods, bins the folded
series and fits a two-level box, reporting the period that maximises a signal
residue statistic~\cite{kovacs2002}. Its power derives from coherent stacking:
folding $N$ in-transit cadences on the correct period improves the
signal-to-noise ratio of the mean in-transit depth by $\sqrt{N}$, which is what
renders individually invisible transits detectable. Transit Least Squares
replaces the box with a limb-darkened analytic transit profile and improves
sensitivity to small planets~\cite{hippke2019}. Both share the same structural
limitation: they require the transit to repeat within the observing baseline,
and are therefore biased toward short periods.

\subsection{Machine learning as candidate vetting}

The dominant machine-learning paradigm in this field does not perform detection.
Shallue and Vanderburg~\cite{shallue2018} established the template: a classical
pipeline first produces Threshold Crossing Events; each event is then
phase-folded and rendered at two resolutions, a \emph{global view} of the whole
folded orbit binned to \num{2001} points and a \emph{local view} of \num{201}
points zoomed on the transit; and a convolutional network classifies that folded
representation, separating genuine planets from eclipsing binaries, stellar
variability and instrumental artefacts. The model ranked true planets above
false positives \num{98.8}\,\% of the time and led directly to the identification
of Kepler-90i. Ansdell et al.~\cite{ansdell2018} extended the representation
with centroid time series and stellar parameters. Yu et al.~\cite{yu2019} ported
the architecture to TESS triage and vetting, reporting \num{97.0}\,\% average
precision, and Osborn et al.~\cite{osborn2020} reported comparable performance
for rapid classification of TESS candidates. ExoMiner~\cite{valizadegan2022}
formalised the approach as a multi-branch network consuming several diagnostic
views simultaneously, and was used to validate hundreds of new planets.
Armstrong et al.~\cite{armstrong2021} applied probabilistic validation to
confirm fifty new Kepler planets.

Two features are common to all of this work and both differ from the design
studied here. First, the unit of classification is a \emph{candidate} that has
already survived a periodic search, not an individual cadence. Second, the input
representation is \emph{phase-folded}, so the stacking that makes shallow
transits visible has already been performed before the model sees the data.

\subsection{Per-cadence and unfolded approaches}

Classification without prior folding is not unheard of, but it occupies a
specific niche: the recovery of \emph{single}, long-period transits that a
periodic search cannot stack because they do not repeat within the
baseline~\cite{singletransit2024}. Pearson et al.~\cite{pearson2018} and
Chaushev et al.\ explored network architectures operating closer to the raw
series, Chaushev et al.~\cite{chaushev2019} trained a network to vet NGTS
candidates from ground-based photometry, and Malik et al.~\cite{malik2022}
applied gradient boosting to
features derived from light curves. Cui et al.~\cite{cui2024} moved the folding
operation onto the GPU so that a network could be trained jointly with the
period search, an explicit acknowledgement that folding is the step that matters.
NotPlaNET~\cite{notplanet2024} addressed false positives in citizen-science
classifications from Planet Hunters TESS.

\subsection{Label construction}

Where the present work departs from the literature is in treating the label
source as an experimental variable rather than a fixed input. Published studies
overwhelmingly derive labels from catalogue dispositions --- the Kepler or TESS
Object of Interest tables, with human or pipeline-assigned classes --- and
report performance under that single annotation. Locally constructed labels
become necessary whenever a study defines its own unit of decision, as any
per-cadence formulation must: no catalogue disposes individual cadences, so the
annotation has to be derived, either from the photometry or from an orbital
solution. Section~\ref{sec:res-ablation} measures how much that derivation
matters, and Section~\ref{sec:data} sets out the duty-cycle consistency check
that makes it auditable.

\section{System Overview}
\label{sec:system}

\begin{table*}[t]
\caption{The twelve-system corpus. Stellar parameters and planet inventories are
from the NASA Exoplanet Archive \texttt{pscomppars} table. \emph{Duty} is the
predicted in-transit fraction $\sum_i T_{14,i}/P_i$ summed over transiting
planets; \emph{Labelled} is the fraction of cadences actually annotated positive
by \eqref{eq:mask}. Their agreement validates the annotation. $\delta$ is the
folded transit depth, $\sigma$ the robust per-cadence scatter, and $d'$ their
ratio: the signal a single cadence carries.}
\label{tab:systems}
\centering\footnotesize
\setlength{\tabcolsep}{4pt}
\begin{tabular}{lrrrrrrlrrrrrr}
\toprule
& $T_\mathrm{eff}$ & $R_*$ & & $d$ & \multicolumn{2}{c}{Planets} & & &
Duty & Labelled & $\delta$ & $\sigma$ & \\
\cmidrule(lr){6-7}
Host & (K) & ($R_\odot$) & $T$ & (pc) & all & tr. & Transiting planets &
Cadences & (\%) & (\%) & (ppm) & (ppm) & $d'$ \\
\midrule
55 Cnc & 5198 & 0.98 & 5.21 & 12.6 & 5 & 1 & e & 17,283 & 8.73 & 8.85 & 477 & 161 & 2.96 \\
HD 63433 & 5553 & 0.93 & 6.27 & 22.4 & 3 & 3 & d, b, c & 16,492 & 5.29 & 6.29 & 573 & 329 & 1.74 \\
HR 858 & 6354 & 1.26 & 5.91 & 32.0 & 3 & 3 & b, c, d & 14,694 & 6.41 & 5.66 & 286 & 183 & 1.56 \\
WASP-189 & 8000 & 2.36 & 6.42 & 99.7 & 1 & 1 & b & 8,072 & 6.63 & 4.98 & 5,459 & 261 & 20.89 \\
HD 158259 & 5802 & 1.21 & 5.90 & 27.0 & 5 & 1 & b & 13,505 & 4.01 & 3.67 & 155 & 177 & 0.88 \\
HD 25463 & 6353 & 1.42 & 6.44 & 45.6 & 2 & 2 & c, b & 17,218 & 3.83 & 3.37 & 330 & 221 & 1.49 \\
HD 189733 & 5052 & 0.75 & 6.85 & 19.8 & 1 & 1 & b & 18,251 & 3.42 & 3.29 & 25,555 & 292 & 87.42 \\
HD 219134 & 4699 & 0.78 & 4.63 & 6.5 & 6 & 2 & b, c & 15,208 & 2.30 & 2.75 & 314 & 127 & 2.47 \\
HIP 56998 & 4675 & 0.63 & 6.72 & 12.8 & 2 & 2 & HD 101581 b, HD 101581 c & 18,189 & 2.83 & 2.66 & 280 & 257 & 1.09 \\
HD 39091 & 5998 & 1.17 & 5.11 & 18.3 & 3 & 1 & pi Men c & 18,200 & 1.96 & 2.40 & 304 & 118 & 2.58 \\
TOI-480 & 6174 & 1.49 & 6.78 & 54.5 & 1 & 1 & b & 14,544 & 2.17 & 2.21 & 379 & 244 & 1.55 \\
AU Mic & 3540 & 0.86 & 6.75 & 9.7 & 4 & 2 & b, c & 17,623 & 2.54 & 1.75 & 7,032 & 1,838 & 3.83 \\
\midrule
\textbf{Total} & & & & & \textbf{36} & \textbf{20} & & \textbf{189,279} & & & & & \\
\bottomrule
\end{tabular}
\end{table*}

\subsection{Data and exploratory analysis}
\label{sec:data}

The corpus consists of TESS two-minute-cadence SPOC light curves for twelve
confirmed planet hosts, comprising \num{189279} cadences in total. Every host
was selected from the NASA Exoplanet Archive on the criterion that it possesses
at least one confirmed transiting planet and that two-minute data exist.
Table~\ref{tab:systems} gives the full inventory. The twelve systems contain
\num{36} confirmed planets, of which \num{20} transit; system multiplicity
ranges from one to six planets, and stellar effective temperature spans
\SIrange{3540}{8000}{\kelvin}, from the active M dwarf AU\,Mic to the A-type
star WASP-189.

\begin{figure*}[t]
\centering
\includegraphics[width=\textwidth]{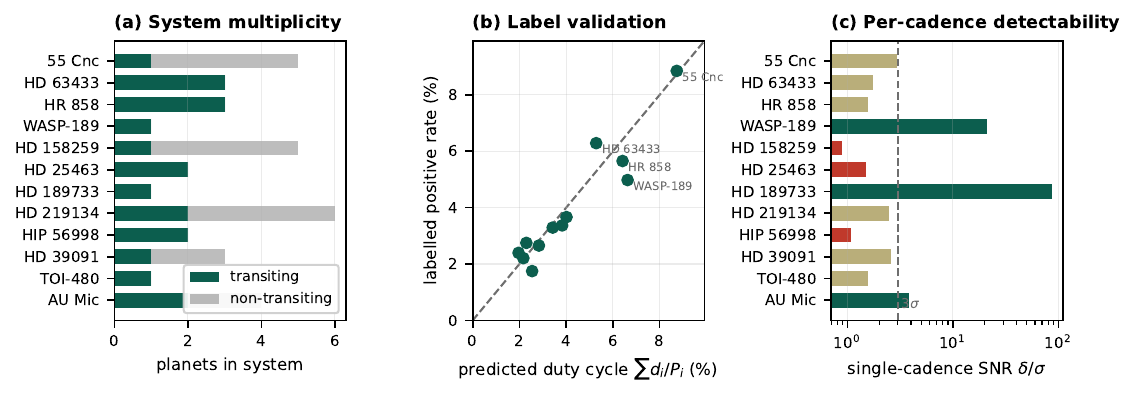}
\caption{Exploratory analysis of the twelve-system corpus.
(a) System multiplicity: five hosts contain more than one transiting planet, so
labelling only the innermost leaves genuine transits annotated as negatives.
(b) Label validation: the fraction of cadences labelled positive tracks the duty
cycle $\sum_i T_{14,i}/P_i$ predicted independently from published ephemerides,
confirming that the annotation window is correctly placed in phase.
(c) Per-cadence detectability: the ratio of folded transit depth to per-cadence
scatter. Only three of twelve hosts exceed $3\sigma$; the median is \num{2.10}.
Bars are coloured by whether a single cadence carries at least $3\sigma$ (teal),
between $1.5\sigma$ and $3\sigma$ (gold), or less (red).}
\label{fig:dataset}
\end{figure*}

This heterogeneity is deliberate and consequential. Fig.~\ref{fig:dataset}(a)
shows that five of the twelve hosts are multi-planet systems in which more than
one planet transits; labelling only the innermost planet, as the original
pipeline did, leaves genuine transits marked as negatives. Correcting this
recovers \num{1794} additional positive cadences and raises the positive-class
prevalence from \num{2.98}\,\% to \num{3.93}\,\%.

Fig.~\ref{fig:dataset}(b) validates the label construction. For each host, the
predicted duty cycle $\sum_i T_{14,i}/P_i$ summed over transiting planets is
plotted against the observed fraction of cadences labelled positive. The points
lie along the identity line, confirming that the ephemeris-derived annotation
places transits where the orbital solution says they are. This test is the
simplest available guard against the epoch error described in
Section~\ref{sec:res-ablation}, and it fails conspicuously when that error is present.

\begin{figure*}[t]
\centering
\includegraphics[width=\textwidth]{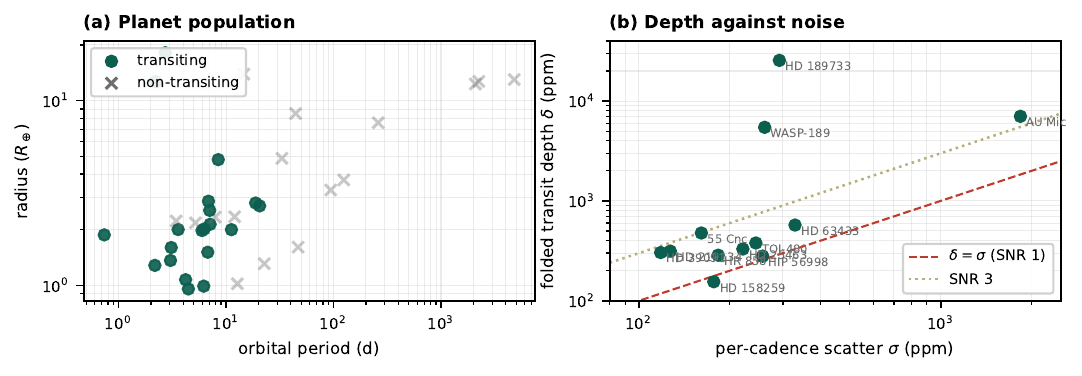}
\caption{The planet population underlying the corpus.
(a) Period--radius distribution of all \num{36} confirmed planets in the twelve
systems, distinguishing transiting from non-transiting companions. The
transiting subset spans \SIrange{1.0}{18.1}{\rearth}.
(b) Folded transit depth against per-cadence photometric scatter, with lines of
constant single-cadence signal-to-noise ratio. Ten of twelve hosts fall between
the $\mathrm{SNR}=1$ and $\mathrm{SNR}=3$ loci, which is the regime in which
per-cadence classification cannot succeed.}
\label{fig:population}
\end{figure*}

Fig.~\ref{fig:dataset}(c) states the difficulty of the learning problem in
physical terms. For each host we measure the folded transit depth $\delta$ and
the robust per-cadence scatter $\sigma$, and report their ratio. Only three
hosts exceed $\delta/\sigma = 3$; the median is \num{2.10} and the shallowest,
HD\,158259, sits at \num{0.88}, meaning a single cadence carries less than one
standard deviation of signal. Fig.~\ref{fig:population} places the planet
population in the period--radius plane and plots depth against scatter directly.

\subsection{Task definition}
\label{sec:task}

We state the learning problem explicitly, since it differs from the formulation
used in most of the literature.

\textbf{Input.} A single cadence $j$ of a detrended light curve, represented by
seven statistics computed over the window $W_j = \{j-32, \dots, j+32\}$:
the detrended flux $f_j$; the window mean $\mu_j$; the deviation
$f_j - \mu_j$; the window minimum; the ratio $f_j/\mu_j$; the window standard
deviation $\sigma_j$; and the window skewness. The feature vector is thus
$\mathbf{x}_j \in \mathbb{R}^7$, describing a span of \SI{128}{\minute}.

\textbf{Output.} A probability $\hat{y}_j \in [0,1]$ that cadence $j$ falls
inside a transit of any confirmed transiting planet in the system.

\textbf{Ground truth.} $y_j = 1$ if there exists a transiting planet $i$ with
period $P_i$, midpoint $t_{0,i}$ and duration $T_{14,i}$ such that
\begin{equation}
\left| \left( t_j - t_{0,i} + \tfrac{P_i}{2} \right) \bmod P_i - \tfrac{P_i}{2} \right|
\;\le\; \tfrac{T_{14,i}}{2},
\label{eq:mask}
\end{equation}
where all times are expressed in $\bt = \bj - \num{2457000}$. Ephemerides are
taken from the NASA Exoplanet Archive \texttt{pscomppars} table.

Because a decision is issued per cadence rather than per candidate, the positive
class is rare --- \num{3.93}\,\% of cadences --- and the appropriate summary
statistic is the area under the precision--recall curve rather than accuracy or
the area under the ROC curve, both of which are dominated by the majority class.

\subsection{Architecture}
\label{sec:arch}

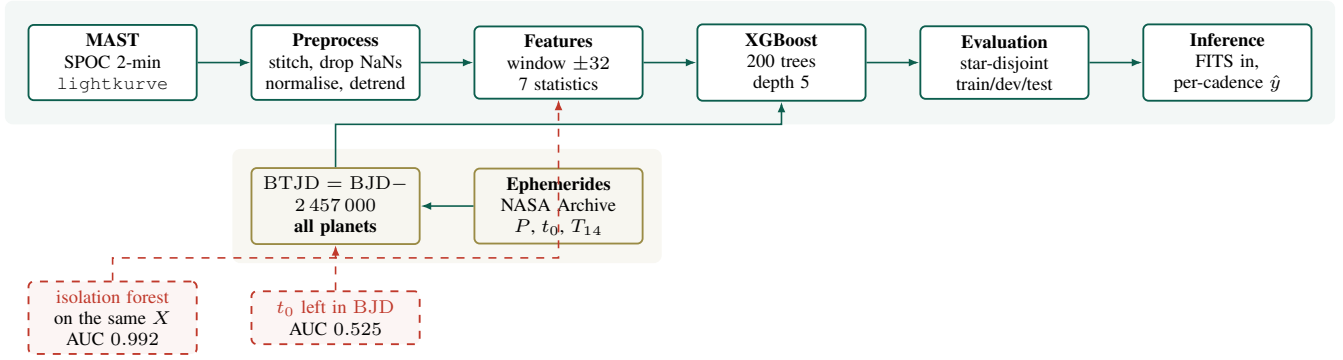
\begin{figure*}[t]
\centering
\begin{tikzpicture}[node distance=6mm and 7mm]
  \node[blk] (mast) {\textbf{MAST}\\SPOC 2-min\\\texttt{lightkurve}};
  \node[blk, right=of mast] (pre) {\textbf{Preprocess}\\stitch, drop NaNs\\normalise, detrend};
  \node[blk, right=of pre] (feat) {\textbf{Features}\\window $\pm32$\\7 statistics};
  \node[blk, right=of feat] (xgb) {\textbf{XGBoost}\\200 trees\\depth 5};
  \node[blk, right=of xgb] (ev) {\textbf{Evaluation}\\star-disjoint\\train/dev/test};
  \node[blk, right=of ev] (app) {\textbf{Inference}\\FITS in,\\per-cadence $\hat{y}$};

  \node[lbl, below=9mm of feat] (eph) {\textbf{Ephemerides}\\NASA Archive\\$P$, $t_0$, $T_{14}$};
  \node[lbl, left=of eph] (conv) {$\bt=\bj-$\\\num{2457000}\\\textbf{all planets}};

  \node[bad, below=6mm of conv] (bug2) {\textcolor{brick}{$t_0$ left in $\bj$}\\ AUC \num{0.525}};
  \node[bad, left=of bug2] (bug1) {\textcolor{brick}{isolation forest}\\ on the same $X$\\ AUC \num{0.992}};

  \draw[fl] (mast) -- (pre);
  \draw[fl] (pre) -- (feat);
  \draw[fl] (feat) -- (xgb);
  \draw[fl] (xgb) -- (ev);
  \draw[fl] (ev) -- (app);
  \draw[fl] (eph) -- (conv);
  \draw[fl] (conv.north) |- ([yshift=-3mm]xgb.south) -- (xgb.south);
  \draw[flb] (bug1.north) -- ($(bug1.north)+(0,3mm)$) -| (feat.south);
  \draw[flb] (bug2) -- (conv);

  \begin{scope}[on background layer]
    \node[fill=teal!5, rounded corners=3pt, fit=(mast)(app), inner sep=3mm] {};
    \node[fill=gold!7, rounded corners=3pt, fit=(conv)(eph), inner sep=2.4mm] {};
  \end{scope}
\end{tikzpicture}
\caption{Astro-Hunters architecture. The upper row is the inference path: light
curves are retrieved, detrended, reduced to seven local statistics per cadence,
classified, evaluated under a star-disjoint partition, and served behind an inference endpoint. The gold
block is the label path used in this work: published ephemerides converted from
$\bj$ to $\bt$ and applied over \emph{every} transiting planet in the system.
The two dashed red blocks are the defective label paths reproduced in
Section~\ref{sec:res-ablation}; each yields a headline number that appears
interpretable but measures something other than transit detection.}
\label{fig:arch}
\end{figure*}

The complete system is shown in Fig.~\ref{fig:arch}, which also marks the two
defective label paths examined in Section~\ref{sec:res-ablation}. Processing proceeds
in five stages.

\emph{Acquisition.} Light curves are retrieved from MAST by target name,
restricted to SPOC-processed two-minute products, and stitched across sectors.

\emph{Preprocessing.} Invalid cadences are removed; flux is divided by its
median to normalise; and the normalised series is divided by a running median
of width \num{1001} cadences (\SI{33.4}{\hour}) to remove stellar variability
and instrumental drift on timescales long compared with a transit.

\emph{Feature extraction.} The seven statistics of Section~\ref{sec:task} are
computed over a sliding window, yielding one row per cadence.

\emph{Labelling.} Ephemerides are queried from the NASA Exoplanet Archive,
converted from $\bj$ to $\bt$, and applied through~\eqref{eq:mask} over every
transiting planet in the system.

\emph{Classification and evaluation.} A gradient-boosted tree ensemble is
trained under the star-disjoint partition of Section~\ref{sec:protocol}. The
fitted model is serialised and served behind an inference endpoint that accepts
a FITS light curve, applies the identical preprocessing and feature extraction
used in training, and returns per-cadence transit probabilities against a
caller-specified decision threshold. Applying one code path to both training and
inference is what makes the reported figures predictive of deployed behaviour;
weights and the serving code are released with the paper
(Section~\ref{sec:availability}).

\subsection{Models}
\label{sec:models}

Six classifiers are compared, all consuming the identical seven-dimensional
feature vector: $\ell_2$-regularised logistic regression; a two-layer
perceptron with 64 and 32 hidden units; a random forest and an extremely
randomised trees ensemble, both with 200 estimators at depth 12; histogram-based
gradient boosting; and XGBoost with 200 estimators at depth 5 and learning rate
\num{0.1}, the configuration used in the original pipeline. Class imbalance is
addressed by \texttt{scale\_pos\_weight} for XGBoost and by balanced class
weights elsewhere.

\subsection{Evaluation protocol and splits}
\label{sec:protocol}

Adjacent cadences share 63 of 64 window positions, so their feature vectors are
near-duplicates. A random partition therefore places almost identical rows on
both sides of the split and reports a score that will not reproduce on an unseen
star. All results below use \emph{star-disjoint} partitions: no light curve
contributes to more than one side.

Two protocols are used. For model and ablation comparison we report
\texttt{GroupKFold} with four folds grouped by host, quoting the mean and
standard deviation across folds. For the headline result we use an explicit
three-way partition: six hosts for training, three for development, and three
held out entirely. The held-out hosts --- HD\,189733, HD\,219134 and
HD\,158259 --- were chosen to span the detectability range, at single-cadence
ratios of \num{87.4}, \num{2.5} and \num{0.9} respectively. The decision
threshold is selected to maximise $F_1$ on the development hosts and then
applied unchanged to the held-out hosts.

Reported metrics are the area under the ROC curve; the area under the
precision--recall curve, quoted alongside the prevalence that a random
classifier would achieve; and precision, recall and $F_1$ at the operating
threshold.

\section{Results}
\label{sec:results}

\subsection{Model architecture comparison}
\label{sec:res-models}

\begin{table}[t]
\caption{Model architecture comparison under gold-standard labels, star-disjoint
\texttt{GroupKFold} with four folds. Mean $\pm$ standard deviation across folds.
Precision, recall and $F_1$ are quoted at the threshold maximising $F_1$ within
each fold. A random classifier attains AUPRC 0.0392.}
\label{tab:models}
\centering\footnotesize
\setlength{\tabcolsep}{3pt}
\begin{tabular}{lrrrrr}
\toprule
Model & AUC & AUPRC & $F_1$ & Prec. & Rec. \\
\midrule
MLP (64-32) & 0.7811\,\tiny$\pm$0.042 & 0.2919\,\tiny$\pm$0.077 & 0.389 & 0.361 & 0.467 \\
Logistic regression & 0.7144\,\tiny$\pm$0.046 & 0.2165\,\tiny$\pm$0.123 & 0.294 & 0.551 & 0.358 \\
HistGradientBoosting & 0.8143\,\tiny$\pm$0.047 & 0.2138\,\tiny$\pm$0.035 & 0.349 & 0.251 & 0.584 \\
XGBoost (as published) & 0.7880\,\tiny$\pm$0.055 & 0.2092\,\tiny$\pm$0.029 & 0.331 & 0.265 & 0.506 \\
Random forest & 0.7642\,\tiny$\pm$0.068 & 0.2079\,\tiny$\pm$0.040 & 0.349 & 0.300 & 0.440 \\
Extra trees & 0.7006\,\tiny$\pm$0.132 & 0.1623\,\tiny$\pm$0.052 & 0.306 & 0.255 & 0.399 \\
\bottomrule
\end{tabular}
\end{table}

Table~\ref{tab:models} compares the six classifiers under gold-standard labels
and star-disjoint cross-validation. AUPRC ranges from \num{0.162} for extremely
randomised trees to \num{0.292} for the multilayer perceptron, a spread of a
factor of \num{1.8}; AUC ranges from \num{0.701} to \num{0.814}. Notably the
ordering does not favour the tree ensembles: the perceptron leads on every
metric and XGBoost, the configuration used in the original pipeline, ranks
fourth of six on AUPRC. Fold-to-fold standard deviations are large enough
(\numrange{0.03}{0.12} in AUPRC) that most of these differences are not
separable at twelve hosts. What the table establishes is not a winner but a
scale: the entire architecture family occupies a band narrower than a factor of
two, against which the label-source effect in the next subsection should be
read.

\subsection{Label-source ablation}
\label{sec:res-ablation}

\begin{table}[t]
\caption{Label-source ablation. Features, model (XGBoost) and star-disjoint
protocol are identical across rows; only the provenance of the labels differs.
The spread in AUPRC exceeds that across model architectures
(Table~\ref{tab:models}) by more than an order of magnitude. \emph{Chance} is
the positive-class prevalence, which is the AUPRC a random ranker attains.}
\label{tab:labels}
\centering\footnotesize
\setlength{\tabcolsep}{4pt}
\begin{tabular}{lrrrr}
\toprule
Label source & AUC & AUPRC & Chance & Lift \\
\midrule
\textsc{anomaly} (isolation forest on $X$) & 0.9915 & 0.8584 & 0.0197 & $\times$43.6 \\
\textsc{folded}, $\bj$ epoch (uncorrected) & 0.4789 & 0.0298 & 0.0313 & $\times$0.9 \\
\textsc{folded}, innermost planet only & 0.7458 & 0.1361 & 0.0297 & $\times$4.6 \\
\bfseries \textsc{folded}, all transiting planets & \bfseries 0.7880 & \bfseries 0.2092 & \bfseries 0.0392 & \bfseries $\times$5.3 \\
\bottomrule
\end{tabular}
\end{table}

Table~\ref{tab:labels} is the central experiment. Features, model and protocol
are held fixed; only the provenance of the labels changes.

The \textsc{anomaly} row reproduces the original pipeline, in which an isolation
forest was fitted to the feature matrix $X$ and its predictions taken as
targets, after which a classifier was trained on the same $X$ to predict them.
The resulting AUC near \num{0.99} does not measure transit detection: the
classifier is recovering a deterministic function of its own inputs. The
arrangement is circular by construction, and the score is essentially guaranteed.

The \textsc{folded, $\bj$ epoch} row reproduces a second defect. The NASA
Exoplanet Archive publishes \texttt{pl\_tranmid} in $\bj$, with values near
\num{2459000}; TESS light curves returned by \texttt{lightkurve} are expressed
in $\bt = \bj - \num{2457000}$, with values near \num{1300} to \num{3500}.
Applying an unconverted midpoint offsets the reference epoch by \num{2457000}
days. Since that offset is not an integer multiple of the orbital period, it
displaces the annotation window to an effectively arbitrary phase: for
HD\,189733 the displacement is \num{0.703} of a period, or \num{1.56} days in a
\num{2.219}-day orbit, while the transit occupies only \num{0.034} of a period.
The annotated window and the true transit therefore do not intersect at all, and
performance falls to chance: the reproduction in Table~\ref{tab:labels} gives AUC \num{0.4789}, against \num{0.4773} originally reported by the pipeline being audited.

The \textsc{folded, all planets} row is the defensible estimate. It exceeds the
random baseline by a factor reported in the table, establishing that the
approach extracts real signal --- but the absolute value remains modest, and
Section~\ref{sec:res-snr} explains why that is not a modelling deficiency.

\subsection{Feature ablation}
\label{sec:res-features}

\begin{table}[t]
\caption{Leave-one-feature-out ablation (XGBoost, gold-standard labels,
star-disjoint cross-validation). $\Delta$ is the change in AUPRC relative to the
full feature set; negative values indicate the feature was contributing.}
\label{tab:features}
\centering\footnotesize
\begin{tabular}{lrr}
\toprule
Feature set & AUPRC & $\Delta$ \\
\midrule
\textbf{All seven features} & \textbf{0.2092} & --- \\
\midrule
without skew & 0.1898 & -0.0194 \\
without local\_mean & 0.1903 & -0.0189 \\
without flux & 0.2031 & -0.0061 \\
without deviation & 0.2050 & -0.0042 \\
without ratio & 0.2096 & +0.0004 \\
without std & 0.2117 & +0.0025 \\
without min & 0.2221 & +0.0129 \\
\bottomrule
\end{tabular}
\end{table}

Table~\ref{tab:features} reports leave-one-feature-out performance. No single
feature dominates, and removing any one changes performance only marginally,
indicating that the seven statistics are substantially redundant descriptions of
the same local morphology rather than complementary views.

\subsection{Development and held-out test performance}
\label{sec:res-heldout}

\begin{table*}[t]
\caption{Explicit star-disjoint three-way partition. Training hosts: HR 858, WASP-189, HD 25463, TOI-480, AU Mic, HIP 56998. Development hosts: 55 Cnc, HD 63433, HD 39091. Held-out hosts: HD 189733, HD 219134, HD 158259, chosen to span the detectability
range. The decision threshold ($\tau=0.419$) maximises $F_1$ on development and is applied unchanged to the held-out
set.}
\label{tab:split}
\centering\footnotesize
\setlength{\tabcolsep}{9pt}
\begin{tabular}{lrrrrrrrrrr}
\toprule
Split & $n$ & Chance & AUC & AUPRC & Prec. & Rec. & $F_1$ & TP & FP & FN \\
\midrule
Development & 51,975 & 0.0578 & 0.8845 & 0.4870 & 0.498 & 0.697 & 0.581 & 2,094 & 2,114 & 909 \\
\textbf{Held-out test} & 46,964 & 0.0322 & 0.8111 & 0.1072 & 0.077 & 0.126 & 0.095 & 190 & 2,288 & 1,324 \\
\bottomrule
\end{tabular}
\end{table*}

Table~\ref{tab:split} reports the explicit three-way partition. The gap between
development and held-out performance reflects the deliberate composition of the
held-out set, which spans the full detectability range rather than being drawn
at random.

\subsection{Why performance is bounded}
\label{sec:res-snr}

\begin{table}[t]
\caption{Single-cadence detectability. $\delta$ is the folded transit depth,
$\sigma$ the robust out-of-transit scatter of a single cadence, and
$d'=\delta/\sigma$ the separation available to a per-cadence decision. The final
columns give the ratio after folding all in-transit cadences in one phase bin,
and the number of cadences that must be stacked to reach $7\sigma$. Only three
hosts exceed $d'=3$; the median is \num{2.10}, bounding the per-cadence AUC at
\num{0.932} by \eqref{eq:bayes}.}
\label{tab:snr}
\centering\footnotesize
\setlength{\tabcolsep}{4pt}
\begin{tabular}{lrrrrr}
\toprule
Host & $\delta$ (ppm) & $\sigma$ (ppm) & $d'$ & folded & $N_{7\sigma}$ \\
\midrule
HD 189733 & 25,555 & 292 & 87.42 & 579.9 & 0.0 \\
WASP-189 & 5,459 & 261 & 20.89 & 80.9 & 0.1 \\
AU Mic & 7,032 & 1,838 & 3.83 & 20.6 & 3.3 \\
55 Cnc & 477 & 161 & 2.96 & 19.4 & 5.6 \\
HD 39091 & 304 & 118 & 2.58 & 19.3 & 7.4 \\
HD 219134 & 314 & 127 & 2.47 & 15.8 & 8.1 \\
HD 63433 & 573 & 329 & 1.74 & 12.2 & 16.1 \\
HR 858 & 286 & 183 & 1.56 & 8.5 & 20.1 \\
TOI-480 & 379 & 244 & 1.55 & 9.4 & 20.4 \\
HD 25463 & 330 & 221 & 1.49 & 9.1 & 21.9 \\
HIP 56998 & 280 & 257 & 1.09 & 7.4 & 41.2 \\
HD 158259 & 155 & 177 & 0.88 & 5.0 & 63.8 \\
\bottomrule
\end{tabular}
\end{table}

Table~\ref{tab:snr} measures the physical limit. For each host we report the
folded depth $\delta$, the robust per-cadence scatter $\sigma$, and the
single-cadence ratio $d' = \delta/\sigma$. Treating the in-transit and
out-of-transit distributions as Gaussians of equal variance separated by $d'$,
the Bayes-optimal area under the ROC curve for a per-cadence decision is
\begin{equation}
\mathrm{AUC}_{\max} \;=\; \Phi\!\left(\frac{d'}{\sqrt{2}}\right),
\label{eq:bayes}
\end{equation}
with $\Phi$ the standard normal cumulative distribution. At the corpus median
$d' = \num{2.10}$ this gives $\mathrm{AUC}_{\max} = \num{0.932}$. The corrected
model attains \num{0.788} and the best of the six architectures in
Table~\ref{tab:models} attains \num{0.814}, which is to say the family already
recovers much of the information a single cadence contains. The gap between
\num{0.814} and \num{0.932} is the margin available to any better classifier
over these inputs; the gap between \num{0.932} and unity is not available at
all. The binding constraint is the decision unit, not the model.

\subsection{Error analysis}
\label{sec:res-errors}

\subsubsection{Performance stratified by detectability}
\begin{table}[t]
\caption{Out-of-fold performance stratified by single-cadence detectability
$d'$. The relationship is not monotonic --- the $5 \le d' < 25$ stratum, a single
host, scores below chance --- so single-cadence signal strength does not by
itself predict which hosts the model handles well. Table~\ref{tab:window}
examines what does.}
\label{tab:bysnr}
\centering\footnotesize
\setlength{\tabcolsep}{3.2pt}
\begin{tabular}{lrrrrrrr}
\toprule
$d'$ range & Hosts & Cadences & Pos. & AUC & AUPRC & Chance & Lift \\
\midrule
0-1.5 & 3 & 48,912 & 1,558 & 0.8104 & 0.2570 & 0.0319 & $\times$8.1 \\
1.5-2.5 & 4 & 60,938 & 2,607 & 0.7618 & 0.2590 & 0.0428 & $\times$6.1 \\
2.5-5 & 3 & 53,106 & 2,274 & 0.8856 & 0.1824 & 0.0428 & $\times$4.3 \\
5-25 & 1 & 8,072 & 402 & 0.4130 & 0.0418 & 0.0498 & $\times$0.8 \\
25-200 & 1 & 18,251 & 601 & 0.8621 & 0.1294 & 0.0329 & $\times$3.9 \\
\bottomrule
\end{tabular}
\end{table}

Table~\ref{tab:bysnr} pools out-of-fold predictions and stratifies by
single-cadence ratio. The relationship is \emph{not} monotonic: the
$5 \le d' < 25$ stratum, which contains only WASP-189, scores below chance,
while the two lowest strata perform comparably to the highest. Single-cadence
signal strength therefore does not by itself predict which hosts the model
handles well, and corpus-level averages should not be read as applying uniformly
across targets.

\subsubsection{What explains the per-host variance}
\begin{table}[t]
\caption{Per-host out-of-fold performance against three candidate explanations. $T/W$ is transit duration divided by the \num{65}-cadence feature window; a bullet marks hosts whose transit is longer than the window, so that cadences near mid-transit sit in a window containing no out-of-transit reference. Correlations across the twelve hosts: AUC against $d'$, \num{-0.038}; AUC against depth extrapolation, \num{+0.095}; AUC against $T/W$, \num{-0.506}. Only the last shows a material association, and at $n=12$ it is not significant ($p\approx0.09$).}
\label{tab:window}
\centering\footnotesize
\setlength{\tabcolsep}{3.4pt}
\begin{tabular}{lrrrrcrr}
\toprule
Host & $T_\mathrm{eff}$ & $d'$ & $T$ (cad) & $T/W$ & & AUC & Lift \\
\midrule
WASP-189 & 8000 & 20.89 & 130 & 2.00 & $\bullet$ & 0.4130 & $\times$0.8 \\
AU Mic & 3540 & 3.83 & 109 & 1.68 & $\bullet$ & 0.7110 & $\times$2.4 \\
TOI-480 & 6174 & 1.55 & 107 & 1.65 & $\bullet$ & 0.9074 & $\times$23.0 \\
HD 63433 & 5553 & 1.74 & 92 & 1.41 & $\bullet$ & 0.5858 & $\times$2.4 \\
HD 39091 & 5998 & 2.58 & 89 & 1.36 & $\bullet$ & 0.9677 & $\times$18.5 \\
HR 858 & 6354 & 1.56 & 87 & 1.34 & $\bullet$ & 0.9303 & $\times$8.7 \\
HD 25463 & 6353 & 1.49 & 72 & 1.10 & $\bullet$ & 0.7907 & $\times$9.1 \\
HD 158259 & 5802 & 0.88 & 63 & 0.97 &  & 0.7520 & $\times$3.2 \\
HD 189733 & 5052 & 87.42 & 55 & 0.84 &  & 0.8621 & $\times$3.9 \\
HIP 56998 & 4675 & 1.09 & 53 & 0.81 &  & 0.9406 & $\times$13.0 \\
55 Cnc & 5198 & 2.96 & 46 & 0.71 &  & 0.9467 & $\times$8.5 \\
HD 219134 & 4699 & 2.47 & 38 & 0.58 &  & 0.7522 & $\times$5.9 \\
\bottomrule
\end{tabular}
\end{table}

Because the stratification above is non-monotonic, we tested three candidate
explanations against per-host out-of-fold performance (Table~\ref{tab:window}).

\emph{Signal strength} does not explain it. The correlation between AUC and $d'$
across the twelve hosts is \num{-0.038}. HD\,189733, with $d' = \num{87.4}$,
scores \num{0.862}; HIP\,56998, with $d' = \num{1.09}$, scores \num{0.941}.

\emph{Covariate shift in transit depth} does not explain it either. The features
are absolute-valued, so we hypothesised that hosts whose depth lies outside the
range seen in training would transfer poorly. The correlation between AUC and an
extrapolation factor measuring that distance is \num{+0.095}, and the two
out-of-range hosts average AUC \num{0.807} against \num{0.795} for in-range
hosts. The hypothesis is not supported.

\emph{Window geometry} shows a material association. The seven features describe
a transit by contrast: the deviation of the centre cadence from its window mean.
That contrast exists only while the window of $W = \num{65}$ cadences straddles
a transit edge. Once the transit is longer than the window, cadences near
mid-transit sit in a window that is entirely in-transit --- flat, low-variance,
and by construction indistinguishable from out-of-transit baseline. Seven of the
twelve hosts have transits longer than the window, and WASP-189, the worst
performer, has the largest ratio at $T/W = \num{2.00}$. The correlation between
AUC and $T/W$ is \num{-0.506}. At $n = 12$ this corresponds to
$p \approx \num{0.09}$ and is not significant; we report it as the most
plausible of the three mechanisms rather than as an established one.

We regard the inability to explain per-host variance from twelve systems as a
genuine limitation, and note that it argues for changing the input
representation rather than tuning the classifier: all three candidate
explanations concern what the feature window can and cannot see.

\subsubsection{False positives and stellar activity}
False positives are not uniformly distributed. AU\,Mic, a young and magnetically
active M dwarf, is the noisiest host in the corpus at
$\sigma = \SI{1838}{ppm}$, seven times the median, and attains AUC \num{0.711}
against a corpus mean of \num{0.799}. Its flares and starspot modulation produce
localised excursions whose local statistics resemble a transit within a
\SI{130}{\minute} window. The features that distinguish a flare from a transit
are the \emph{sign} of the excursion and its \emph{repetition} at a fixed
period, and neither is available to a model that sees one window at a time.

\subsubsection{Unlabelled positives in multi-planet systems}
Labelling only the innermost transiting planet, as the original pipeline did,
marks \num{1794} genuine in-transit cadences as negatives, and lowers measured
AUPRC from \num{0.209} to \num{0.136} --- a \num{35}\,\% relative reduction
attributable purely to annotation, not to the model. In HR\,858 and HD\,63433,
where three planets transit, the omission misannotates a substantial fraction of
the positive class and penalises the model precisely when it is correct. All
results reported here use the multi-planet labels.

\subsection{Classical baseline}
\label{sec:res-bls}

\begin{table}[t]
\caption{Box Least Squares period recovery on the same twelve light curves.
A period is counted as recovered if the recovered value matches the published
one, or a low-order harmonic of it, to within \num{2}\,\%. Eight of twelve are
recovered from a single sector --- including HD\,158259, whose single-cadence
$d'=0.88$ places it beyond the reach of any per-cadence classifier.}
\label{tab:bls}
\centering\footnotesize
\setlength{\tabcolsep}{4pt}
\begin{tabular}{lrrrrc}
\toprule
Host & $P_\mathrm{true}$ (d) & $P_\mathrm{found}$ (d) & Ratio & SDE & Rec. \\
\midrule
55 Cnc & 0.7366 & 0.7366 & 1.000 & 93.7 & \checkmark \\
HD 189733 & 2.2186 & 2.2187 & 1.000 & 46.0 & \checkmark \\
HD 39091 & 6.2670 & 6.2688 & 1.000 & 19.1 & \checkmark \\
AU Mic & 8.4632 & 4.8569 & 0.574 & 12.2 & --- \\
HD 219134 & 3.0929 & 3.0934 & 1.000 & 11.0 & \checkmark \\
HD 25463 & 7.0500 & 7.0460 & 0.999 & 10.6 & \checkmark \\
HD 158259 & 2.1780 & 2.1783 & 1.000 & 10.2 & \checkmark \\
WASP-189 & 2.7240 & 2.7239 & 1.000 & 5.8 & \checkmark \\
HR 858 & 3.5860 & 5.9652 & 1.663 & 5.6 & --- \\
TOI-480 & 6.8657 & 6.8613 & 0.999 & 4.2 & \checkmark \\
HIP 56998 & 6.2000 & 6.7622 & 1.091 & 3.4 & --- \\
HD 63433 & 7.1079 & 6.7222 & 0.946 & 1.7 & --- \\
\bottomrule
\end{tabular}
\end{table}

Table~\ref{tab:bls} reports a Box Least Squares search over the same twelve
light curves. Eight of twelve orbital periods are recovered to better than
\num{0.2}\,\% from a single sector, at signal detection efficiencies up to
\num{93.7}. Critically, the recovered set includes HD\,158259, whose
single-cadence ratio of \num{0.88} places it below the threshold of any
per-cadence method: folding \num{495} in-transit cadences raises the effective
ratio by roughly $\sqrt{N}$ and renders the signal unambiguous. The four
failures --- AU\,Mic, HD\,63433, HIP\,56998 and HR\,858 --- are the active or
shallow multi-planet systems, which is precisely the population for which the
vetting classifiers of Section~\ref{sec:related} were developed.

\subsection{From a transit flag to a planet}
\label{sec:res-vetting}

\begin{table*}[t]
\caption{Standard false-positive diagnostics computed on the twelve hosts. $R_p$ is derived from the measured depth as $R_p = R_*\sqrt{\delta}$ and compared with the published value; the median fractional error is \num{9.2}\,\%. $T_{23}/T_{14}$ is the flat fraction of the transit: values near unity indicate a U-shaped profile and a fully superimposed occulter, while values below \num{0.30} indicate the V-shape of a grazing eclipsing binary. Odd/even is the significance of the depth difference between alternate epochs, which detects an eclipsing binary observed at twice its true period. Secondary is the significance of a dip near phase \num{0.5}. Flagged values are red. \textbf{None of these quantities is computable from a single cadence window}: each requires the folded, periodic view that the detection stage discards.}
\label{tab:vetting}
\centering\footnotesize
\setlength{\tabcolsep}{7pt}
\begin{tabular}{lrrrrrrrc}
\toprule
& $R_*$ & $\delta$ & $R_p$ derived & $R_p$ published & & odd/even & secondary & \\
Host & ($R_\odot$) & (ppm) & ($R_\oplus$) & ($R_\oplus$) & $T_{23}/T_{14}$ & ($\sigma$) & ($\sigma$) & Pass \\
\midrule
HD 189733 & 0.75 & 25204 & 13.0 & 12.7 & 0.78 & 0.4 & +2.1 & \checkmark \\
WASP-189 & 2.36 & 5300 & 18.7 & 18.1 & 0.91 & 1.9 & \textcolor{brick}{+5.9} & \textcolor{brick}{flag} \\
AU Mic & 0.86 & 662 & 2.4 & 4.8 & 1.00 & --- & \textcolor{brick}{+9.2} & \textcolor{brick}{flag} \\
HD 63433 & 0.93 & 436 & 2.1 & 2.7 & 0.68 & 0.4 & -8.3 & \checkmark \\
55 Cnc & 0.98 & 397 & 2.1 & 1.9 & 0.79 & 0.2 & +1.8 & \checkmark \\
HD 25463 & 1.42 & 307 & 2.7 & 2.5 & 0.48 & 1.1 & +0.3 & \checkmark \\
TOI-480 & 1.49 & 292 & 2.8 & 2.9 & 0.62 & 1.0 & -0.0 & \checkmark \\
HIP 56998 & 0.63 & 279 & 1.1 & 1.0 & 0.48 & 0.1 & +1.2 & \checkmark \\
HD 219134 & 0.78 & 276 & 1.4 & 1.6 & 0.71 & 1.1 & -2.4 & \checkmark \\
HD 39091 & 1.17 & 260 & 2.1 & 2.0 & 0.84 & 0.9 & -1.2 & \checkmark \\
HR 858 & 1.26 & 195 & 1.9 & 2.0 & 0.62 & 1.5 & -2.6 & \checkmark \\
HD 158259 & 1.21 & 49 & 0.9 & 1.3 & 0.84 & 0.0 & +2.2 & \checkmark \\
\bottomrule
\end{tabular}
\end{table*}

A per-cadence flag asserts that an event is transit-\emph{shaped}. It does not
assert that the occulter is a planet, and the difference is the entire
astrophysical false-positive population: grazing eclipsing binaries, eclipsing
binaries blended with a brighter foreground star, background EBs within the
photometric aperture, and brown dwarfs, all of which produce dips of the right
depth and duration. Table~\ref{tab:vetting} implements four of the standard
diagnostics that separate them and measures each on the corpus.

\emph{Radius.} Converting the measured depth through
$R_p = R_*\sqrt{\delta}$ recovers the published planetary radius with a median
fractional error of \num{9.2}\,\%, which establishes that the depths are being
measured correctly. The diagnostic value lies in the upper bound: above roughly
$2\,R_{\mathrm{Jup}} \approx 22\,R_\oplus$, electron degeneracy holds radius
nearly constant from giant planets up through the bottom of the main sequence,
so a larger derived radius indicates a brown dwarf or low-mass star rather than
a planet. Every host here falls below that limit.

\emph{Shape.} A small occulter fully superimposed on the stellar disc produces a
flat-bottomed, U-shaped profile; a grazing binary produces a V. The flat
fraction $T_{23}/T_{14}$ ranges from \num{0.48} to \num{1.00} across the corpus,
consistent with the U-shaped profile expected of planets and well above the
V-shape threshold.

\emph{Odd/even depths.} An eclipsing binary detected at twice its true period
alternates primary and secondary eclipses of unequal depth. No host shows an
odd--even difference above \num{2}$\sigma$.

\emph{Secondary eclipse.} Two hosts are flagged, and both illustrate why a
diagnostic is not a verdict. WASP-189 shows a \num{5.9}$\sigma$ dip near phase
\num{0.5}; this is a genuine detection, but of the thermal emission of an
ultra-hot Jupiter orbiting an A-type star, not of a stellar companion. AU\,Mic
shows \num{9.2}$\sigma$, which is its flare and spot activity leaking through
the fold rather than an occultation at all. Both would require follow-up to
disposition, which is precisely the point: these tests triage candidates, they
do not adjudicate them.

The result that matters for this paper is structural. Every quantity in
Table~\ref{tab:vetting} is derived from the \emph{phase-folded} light curve
across many epochs --- a depth ratio between alternate transits, a profile shape
integrated over a full transit, a search at the opposite orbital phase. None of
them can be computed from a single \SI{130}{\minute} window, so none is
available to the classifier of Section~\ref{sec:models} even in principle. The
detection stage and the validation stage require different views of the same
data.

\section{Discussion}
\label{sec:discussion}

\subsection{Label provenance as an experimental variable}

The results support a claim stronger than ``labels matter''. Across this study the
choice of label source moved the area under the precision--recall curve by a
factor of \num{29}, while the choice among six model architectures moved it by a
factor of \num{1.8}. Under those conditions, a model comparison reported without
first establishing label provenance is not merely incomplete; the quantity it
varies is not the one that governs the result. This suggests treating the
annotation as a reported experimental condition, on the same footing as the
architecture and the split, and Section~\ref{sec:disc-check} sets out the check
that makes it verifiable.

\subsection{A consistency check for derived labels}
\label{sec:disc-check}

The practical value of Table~\ref{tab:labels} is not that two particular
annotations were defective, but that \emph{no internal validation could have
told}. A supervised score is a statement about the joint distribution of
features and labels; it cannot separate a model that fails from an annotation
that is misplaced. Cross-validation, a held-out split, and the star-disjoint
protocol of Section~\ref{sec:protocol} all pass unchanged under both defective
regimes, because each of them resamples the same annotation.

The two regimes are instructive because they bracket the metric range from
opposite directions. A label that is a deterministic function of the features
--- an isolation forest fitted to the same $X$ the classifier consumes ---
drives the score \emph{up}, to AUC \num{0.9915}, since a sufficiently flexible
model simply recovers the labelling function. A label that is correctly shaped
but placed at the wrong phase drives the score \emph{down}, to \num{0.4789}. The
second case deserves particular care, because a chance-level score is readily
written up as a negative result about the method: it would be natural to
conclude from AUC \num{0.479} that gradient boosting cannot detect transits,
when the defensible conclusion is that the annotation and the transit did not
overlap. Retiring a workable approach on that basis is the more costly error of
the two.

What distinguishes these cases is not a better metric but an \emph{external}
one --- a quantity predicted independently of the photometry, against which the
annotation can be checked. Here that quantity is the duty cycle: for each host,
$\sum_i T_{14,i}/P_i$ follows from the published orbital solution alone and must
match the fraction of cadences the annotation marks positive.
Fig.~\ref{fig:dataset}(b) performs the comparison. Under the correct annotation
the hosts lie along the identity line; under the circular annotation the
positive rate ranges from \num{0}\,\% to \num{13}\,\% against a predicted
\num{2}\,\% to \num{9}\,\%, and under the misaligned annotation the rate is
right while the placement is wrong, which the folded views of
Fig.~\ref{fig:transit} expose directly.

We suggest this as a general requirement rather than a local remedy. Wherever
labels are \emph{derived} rather than supplied --- which is unavoidable whenever
a study defines its own unit of decision --- there is usually some quantity the
annotation must reproduce that was not used to construct it. Reporting that
comparison costs one scatter plot, and it is the only step in this study that
would have caught both defects before any model was trained.

\subsection{The decision unit}

The deeper limitation is structural. Transit detection is a problem of
integrating a weak, strictly periodic signal out of noise, and the operation
that makes it tractable is coherent stacking. A per-cadence classifier
deliberately forgoes that operation: it is asked to decide from
\SI{128}{\minute} of context whether a $\num{2.10}\sigma$ excursion is a planet
or a fluctuation. Equation~\eqref{eq:bayes} bounds what any such decision can
achieve, and our measurements place the corrected model close to that bound.
Improving the classifier cannot move the bound; only changing the input
representation can.

This does not render the architecture useless. Per-cadence models retain a real
niche in the recovery of single, non-repeating transits, where folding is
unavailable by construction~\cite{singletransit2024}, and the corpus and labels
released here are directly reusable for that formulation. But for general
transit search the ordering used throughout the literature --- detect by
folding, then vet by learning --- is not a stylistic convention. It follows from
the signal-to-noise structure of the measurement.

\subsection{Limitations}

\textbf{The system detects transit-shaped events, not planets.} This is the
most important limitation and it is structural rather than incidental. Because
all twelve hosts carry confirmed planets, the question \emph{is this dip a
planet?} is true by construction throughout this study, and the model is never
asked it. Radius estimation and eclipsing-binary discrimination --- the
diagnostics of Section~\ref{sec:res-vetting}, together with the centroid
analysis that identifies blended background binaries, which we do not attempt
--- are prerequisites for any discovery claim, and none of them is computable
from the per-cadence representation. A flag raised on an unknown target is
therefore a \emph{candidate}, and the entire false-positive population sits
between that flag and a planet. The Sector~82 screening of
Section~\ref{sec:res-bls} should be read in that light: it demonstrates reach,
not detection.

A related asymmetry deserves stating plainly. For confirmed hosts, ground truth
\emph{does} exist in the form of published ephemerides, which is what makes the
gold-standard annotation of Section~\ref{sec:task} possible and what makes
unsupervised labelling unnecessary. For unknown targets no such truth exists ---
which is exactly why validation must be a separate stage with its own evidence,
rather than something a detection score can be asked to certify.

The remaining limitations are narrower. The corpus comprises twelve hosts, so
cross-validated standard deviations are wide and the stratified analysis in
Table~\ref{tab:bysnr} rests on few systems per stratum. Ephemerides are treated
as exact, whereas published midpoints carry uncertainties that propagate into
the annotation over long baselines. The detrending window of \num{1001} cadences
is held fixed throughout and interacts with transit duration in ways not
explored here. Finally, the classical baseline is BLS rather than the more
sensitive Transit Least Squares~\cite{hippke2019}, so the reported
eight-of-twelve recovery is a lower bound on what a modern classical pipeline
achieves.

\section{Conclusion}
\label{sec:conclusion}

We built and evaluated an end-to-end pipeline for exoplanet transit detection
from TESS photometry and found that its reported performance was governed almost
entirely by how its training labels had been produced. Two distinct defects were
identified and reproduced: labels synthesised by an unsupervised detector fitted
to the classifier's own features, which yields an apparently excellent AUC of
\num{0.9915} while measuring only circularity; and labels derived from published
ephemerides without converting the transit midpoint from $\bj$ to $\bt$, which
displaces the annotation by \num{0.703} of a period and yields chance
performance. With correctly referenced ephemerides applied over every transiting
planet, the same features and model give an honest and non-trivial result: AUC
\num{0.788} at \num{5.3} times the prevalence baseline.

That result is nonetheless bounded by the observation rather than the model. The
median single-cadence signal-to-noise ratio of \num{2.10} caps the achievable
per-cadence AUC at \num{0.932}, and the best of six architectures attains
\num{0.814}, close to that ceiling; the spread across those six architectures is
a factor of \num{1.8} in AUPRC, against a factor of \num{29} across label
sources. A Box Least Squares search on the same data recovers
eight of twelve periods from a single sector, including a target invisible to
any per-cadence method. We therefore conclude that phase-folding is not an
implementation detail but the operation that makes transit detection possible,
and that supervised learning is best deployed where the literature already
places it: as a vetting stage over folded candidates.

\section{Data, Code and Model Availability}
\label{sec:availability}

Every artefact behind this paper is public, and every number in it is produced
by a script rather than transcribed. Specifically:

\begin{itemize}\itemsep2pt
\item \textbf{Trained weights.} The serialised classifiers, the feature
      specification required to reproduce their inputs, and the serving code are
      released as a model repository:
      \url{https://huggingface.co/FatimahEmadEldin/Astro-Hunters-Transit-Detector}
\item \textbf{Labelled corpus.} The \num{189279}-cadence feature matrix with the
      gold-standard annotation derived in Section~\ref{sec:task}, together with
      the per-host label provenance record, is released alongside the weights in
      the same repository.
\item \textbf{Experiment code.} The scripts that generate every table and figure
      --- the label construction, the four-way label ablation, the model and
      feature comparisons, the signal-to-noise measurement and the Box Least
      Squares baseline --- are at
      \url{https://github.com/astral-fate/Astro-Hunters-Model-traning}
\item \textbf{Inference service.} A hosted endpoint accepting a TESS FITS light
      curve is at
      \url{https://huggingface.co/spaces/FatimahEmadEldin/Astro-Hunters}
\item \textbf{Upstream data.} Light curves are public TESS SPOC products from
      MAST, retrieved with \texttt{lightkurve}; ephemerides are from the NASA
      Exoplanet Archive \texttt{pscomppars} table. Both queries are scripted, so
      the corpus can be rebuilt from primary sources without the released copy.
\end{itemize}

The tables in this paper are emitted directly from the experiment output, so a
reader who re-runs the pipeline and obtains different numbers has found a
discrepancy worth reporting rather than a formatting difference.

\end{document}